\documentstyle[12pt]{article}
\newcommand{\be}{\begin{equation}}
\newcommand{\bea}{\begin{eqnarray}}
\newcommand{\eea}{\end{eqnarray}}
\newcommand{\ba}{\begin{array}}
\newcommand{\ea}{\end{array}}
\newcommand{\ee}{\end{equation}}

\expandafter\ifx\csname mathbbm\endcsname\relax

\else

\fi
\begin{document}

\begin{titlepage}
\hfill
\vbox{
    \halign{#\hfil         \cr
           CERN-TH/2000-061 \cr
           hep-th/0002198   \cr
           } 
      }  
\vspace*{20mm}
\begin{center}
{\Large {\bf  On Type II NS5-branes in the presence of an RR field}\\} 

\vspace*{15mm}
\vspace*{1mm}
Mohsen Alishahiha
\vspace*{1cm} 

{\it Theory Division, CERN \\
CH-1211, Geneva, 23, Switzerland}\\

\vspace*{.5cm}
\end{center}

\begin{abstract}
In this letter we use Type II NS5-branes in the presence of an RR field 
at the decoupling limit to study a non-commutative version of little 
string theory. We shall see that the decoupling limit of NS5-branes in 
the presence of an RR field  is completely different from the one we had in 
ordinary little string theory; in particular for Type IIA NS5-branes this 
decoupling limit can be defined only when the theory is wrapped on a circle,
but nevertheless flows to the ordinary non-compact (0,2) theory at the 
IR limit. We also see that these theories, in the UV regime, where the 
non-commutative effects are important, can be described by smeared
ordinary D3(D2)-branes in Type IIB(A) string theory.

\end{abstract}
\vskip 4cm

February 2000
\end{titlepage}

\newpage

\section{Introduction}

It was shown in \cite{DH} (see also \cite{JA}) that, when we turn on a B field on the D-brane
worldvolume, the low-energy effective worldvolume theory is modified to be a
non-commutative Super-Yang-Mills (NCSYM) theory. In fact the worldvolume theory
of $N$ coincident D$_p$-branes in the presence of a B field is found
to be $U(N)$ NCSYM theory \cite{SW}.

As is the case with a zero B field, there exists a limit where the bulk
modes decouple from the worldvolume non-commutative field theory
\cite{{SW},{EKMNW}}; we expect to have a correspondence between string theory
on the curved background with B field and non-commutative field theories.
In other words we expect to have a holographic picture like AdS/CFT 
correspondence \cite{M} (for review and complete list of references, 
see \cite{AGMOO}) for the non-commutative theories. In fact this issue has
been investigated in \cite{HI}-\cite{HO}.

It was also shown that there is a limit in which the theory living
 on the NS5-branes decouples from the
bulk \cite{SI1}\footnote{ This idea, that the theory on the NS5-branes could be
some sort of string theory, has been also considered in \cite{{BRS},{DVV}} in the
context of Matrix theory description of M-theory and its compactifications.}.
It is believed that the decoupled theory is a string theory
without gravity called ``{\it Little String Theory}'' (LST) \footnote{
For a brief review, see \cite{OFER}.}. Upon
compactification this theory inherits the T-duality of Type II string theories
and must therefore be a non-local theory. The simplest way to find this decoupling
limit is to start with the decoupling of D5-branes and using S-duality. 
The decoupling limit of D5-branes is defined by $l_s\rightarrow 0$ 
keeping $g^2_{YM}=g_sl_s^2$ fixed. Using S-duality \footnote{Under S-duality we
have $l_s^2\rightarrow {l'}_s^2 \equiv g_sl_s^2$ and $g_s \rightarrow g'_s
\equiv\frac{1}{g_s}$} one finds
\be  
g'_s\rightarrow 0\;\;\;\;\;\;\;\;\; g^2_{YM}={l'}^2_s={\rm fixed}
\label{OIIB}
\ee
which is the decoupling limit of NS5-branes \cite{SI1}. Using T-duality we can
also obtain the limit in which the Type IIA NS5-branes decouple from the bulk,
which is the same as the IIB one (\ref{OIIB}).

The NS5-branes theory, or LST, has been studied from several points of view; in
particular it has been considered in \cite{ABKS} using the holographic
principle for string theory in an asymptotic linear dilaton background.
There, the authors considered the string theory in the near-horizon background
of parallel NS5-branes, which is dual to LST. This duality could help them
to study some observables of the theory as well as some of their correlation
functions. The theory on $N$ IIA NS5-branes has also been studied at large
$N$ using supergravity in \cite{MS}. 

It is a natural question to ask what the non-commutative deformation
of LST is. In the spirit of what we have learnt about D-branes in the presence
of a B field and its relation with non-commutative gauge theory due to holography
(AdS/CFT correspondence), we would expect that the non-commutative deformation 
of LST could be obtained by NS5-branes in the presence of a non-zero
RR field along its worldvolume. In fact such a background has been considered in
\cite{AJO}, and it is the aim of this letter to study this theory in more
detail. 

The non-commutative deformation of LST is also expected from the DLCQ description
of LST \cite{AB}; there the authors showed that blowing up the singularities 
by FI terms can give, in the DLCQ context, a space-time interpretation that
could be considered as turning on an RR field parallel to the NS5-branes. 
Therefore we would expect to find a non-commutative version of LST (NLST).

In this letter we study the theory living on the NS5-branes in the
presence of an RR field at its  decoupling limit. We shall only
consider the case with the smallest rank of non-zero RR field. 
This theory can be considered as a non-commutative deformation of little string 
theory, although as we shall see the decoupling limits of non-commutative and
ordinary LST are completely different. In particular in this case the 
decoupling limit is defined as a limit where the string scale goes to zero. Although
even in this limit we have a scale in the theory (coming from the RR field), 
it is not clear whether the theory on the``{\it non-commutative NS5-branes}'' 
obtained in this way is a
string theory. In fact, as we shall see, the theory seems to be equivalent to 
the SYM theory. This may be because of our definition of the decoupling limit
of the theory, but till now, it is the only consistent way this decoupling 
limit can be defined. Nevertheless this theory has some properties
of LST; for example we would expect that the thermodynamical quantities
of both theories be the same, both theories are non-local and have 
some sort of T-duality. It might also be possible that what we are studying here
is not really a non-commutative deformation of LST. In other words, what we have
obtained by S- or T-duality as a decoupling limit of the NS5-branes would be
a limit that leads to a non-commutative field theory, such as SYM theory on the
NS5-branes with an RR field, and in fact it might be that the non-commutative
little string theory does not have a supergravity description\footnote{This
point was suggested by Y. Oz.}. 

In section 2 we shall consider the Type IIB NS5-branes in the presence of
an RR field and we will see that in the UV limit where non-commutative effects are
important the theory can be described by smeared D3-branes. In section 3 we
will study Type IIA NS5-branes. We find that the decoupling limit of the theory
is consistent only when the theory is wrapped on a circle and this theory
can be described at UV by smeared D2-branes. We will give
a conclusion and some comments in section 4.

\section{Non-commutative Type IIB NS5-branes}

{\large {\it Decoupling limit}}

As for the ordinary NS5-brane, where there was a non-trivial theory on its worldvolume
in its decoupling limit, we would expect to find a non-commutative
version of little string theory on the worldvolume of the NS5-brane
in the presence of a non-zero RR field in its decoupling limit. 
The simplest way to define this theory is to start with a D5-brane in Type IIB
string theory in the presence of a non-zero B field along its worldvolume and 
using S-duality to map the theory to its S-dual. Doing so we end up with the
NS5-brane solution with an RR field background.

The decoupling limit of D5-branes in the presence of a large B field is
defined as a limit where $l_s\rightarrow 0$ while $g_s$ and
$b=l_s^2 B$ are kept fixed; moreover, we have to rescale the directions in which
the B field is defined, thereby making them non-commutative.
In this limit the modes on the D5-brane decouple from modes on the bulk and
we are left with 6-dimensional non-commutative SYM with gauge coupling
$g^2_{YM}\sim g_s b$. The phase diagram of this theory has been studied in
\cite{AJO}.

Using S-duality we can find a decoupling limit in which we expect
to have a decoupled theory on the worldvolume of the NS5-brane in the presence
of an RR field: 
\be
l'_s\rightarrow 0,\;\;\;\;\; b'=g'_s{l'}^2_sA={\rm fixed},\;\;\;\;\; g'_s=
{\rm fixed} 
\ee
where $A$ is the RR field obtained from a B field by S-duality.
The coupling of the theory will be $g_{YM}^2=\frac{b}{g'_s}=b'$.

We note that the decoupling limit of the NS5-brane with the RR field is completely
different from ordinary NS5-brane (\ref{OIIB}). In fact, in both the ordinary
case and
the case with an RR field, what we want to send to zero is $g'_s{l'}_s^2$, but the
important point is which quantity we would like to keep fixed. Since we want
to have a non-trivial theory at the decoupling limit, it is natural to assume 
that the coupling of the theory should be fixed. In general for an RR field with 
rank $2m$, the coupling is proportional to $g_s^{-m}l_s^{2-2m}$. 
Now we can see that adding a non-zero RR field will change the decoupling 
limit in a completely non-trivial way.  

We also note that the decoupling limit of NLST looks very much like 
ordinary D3-branes. We will go back to this point later.

\vspace*{.4cm}

{\large {\it Supergravity description}}

\vspace*{.4cm}

We can study $N$ coincident NS5-branes in the presence of an RR field at 
large $N$ 
using supergravity. In order to study this theory one can start with 
the supergravity solution of D5-branes with a B field and using S-duality.
The supergravity solution of D5-branes in the presence of a rank-two B field 
is given by:
\bea
ds^2&=&f^{-1/2}[-dx_{0}^2+\cdots+h(dx_4^2+dx_{5}^2)
]+f^{1/2}(dr^2+r^2d\Omega_{3}^2),\cr
&&\cr
f&=&1+\frac{Ng_sl^2}{\cos\theta\; r^{2}},\;\;\;\;\;\;
h^{-1}=\sin^2 \theta f^{-1}+\cos^2 \theta,\cr
&&\cr
B_{45}&=&\frac{\sin \theta}{\cos \theta}f^{-1}h,\;\;\;\;\;\;\;\;\;
e^{2\phi}=g_s^2f^{-1} h.
\label{SG}
\eea

Under S-duality we have
\be
e^{\phi}\rightarrow e^{\phi'}\equiv e^{-\phi},\;\;\;\;\;
ds^2\rightarrow ds'^2\equiv g_se^{-\phi}ds^2 \ .
\label{st}
\ee
Using (\ref{st}) we get the Type IIB NS5-branes background \cite{AJO}:
\bea
{ds'}^2&=&h^{-1/2}[-dx_{0}^2+\cdots+h(dx_4^2+dx_{5}^2)
+f(dr^2+r^2d\Omega_{3}^2)],\cr
&&\cr
f&=&1+\frac{ N {l'}_s^2}{\cos \theta\; r^2},\;\;\;\;\
h^{-1}=\sin^2 \theta f^{-1}+\cos^2 \theta,\cr
&&\cr
e^{2\phi'}&=&{g'}_s^2f h^{-1} \ ,
\label{NSS}
\eea
and the NS field $B_{ij}$ is mapped to the RR field $A_{ij}$.
The decoupling limit is derived by applying S-duality on
the decoupling limit of the D5-branes.
As we said above, it is defined by taking the limit ${l'}_s^2 \rightarrow 0$ 
and keeping fixed
\be\ba {ll}
u=\frac{r}{{l'}_s^2}&{\bar g'}_s={g'}_s^{-1} \,\cr
b'={l'}_s^{2}\tan \theta &
{\bar x}_{4,5}=\frac{b'}{{l'}_s^2}x_{4,5} \ .
\label{iib}
\ea\ee
Keeping $u$ fixed means keeping fixed the mass of a D-string stretched 
between two NS5-branes. In this limit the supergravity solution reads
\bea
{ds'}^2&=&\frac{{l'}_s^2}{b'}h^{1/2}\left[-dx_{0}^2+\cdots+
h^{-1}(dx_4^2+dx_{5}^2)
+\frac{Nb'}{u^2}(du^2+u^2d\Omega_{3}^2)\right],\cr
&&\cr
h&=&1+(au)^2,\;\;\;\;\;\;\;\;\;\; a^2=\frac{b'}{N},\;\;\;\;\;\;\;\;\;\;
e^{2\phi'}=g_s^{\prime 2} \frac{1+(au)^2}{(au)^2}\ .
\label{NSBD}
\eea
The curvature of the metric reads
\be
{l'}_s^2{\cal R}\sim \frac{1}{N}\;\frac{1}{(1+(au)^2)^{1/2}} \ .
\label{RNS}
\ee
When $au \gg 1$, which we are interested in, the supergravity approximation 
can be trusted for finite $N$, which means that we can study NLST even for 
finite $N$. In this limit the background reads
\bea
{l'}^2_s{ds'}^2&=&\frac{u}{R^2}(-dx_0^2+\cdots+dx_3^2)+\frac{R^2}{u}
(du^2+u^2d\Omega_3^2)+\frac{R^2}{{b'}^2u}(dx_4^2+dx_5^2)\cr
e^{2\phi'}&=&{\bar g}_s^{\prime 2},\;\;\;\;\;\;\; R^2=\sqrt{Nb'},
\eea
and the curvature of the metric is ${l'}^2_s{\cal R}\sim\frac{1}{R^2 u}$.
From the dilaton we can see that this solution is similar to D3-branes and
it is in fact the 
D3-branes solution smeared in two directions and without RR field.

The same situation has also been studied for D$_p$-branes in \cite{{LR},{CO}}.
 There, 
it was shown that the non-commutative $(p+1)$-dimensional SYM theory
can be considered as $(p-1)$-dimensional ordinary YM theory whose gauge group
is obtained by the B field. In fact what the authors have shown is as follows:
they observed that for D$_p$-branes in the presence of a B field with rank two 
at UV, where the non-commutative effects are important, the
supergravity solution reduces to D$_{(p-2)}$-branes smeared in two
directions without B field. In fact this was the case because in this limit
the physics is described by D$_{(p-2)}$-branes; since in this case the B field
that we started with is not along the world-volume of the D$_{(p-2)}$-branes, it
can be gauged away, and we end up with an ordinary smeared brane.

In our example the situation is the same as D5-branes where the theory can
be described by smeared D3-branes, but in our case we have to gauge away the
RR field instead of the B field, which is made possible by the $SL(2,Z)$ symmetry
of Type IIB string theory. We also note that the smeared D3-brane is also
self-dual under S-duality. The reason is that, in smeared D3-branes,
only the harmonic function of the metric will change and the
dilaton and 4-form field will be the same as localized D3-branes. Therefore 
we would expect that the $SL(2,Z)$ symmetry maps the smeared D3-branes to itself.

Since both the non-commutative D5-branes theory and the non-commutative
NS5-branes can be described by ordinary D3-branes smeared in two directions, and
moreover, that this D3-brane solution is self-dual under
S-duality, we conclude that these two theories must be the same.    

For $au\ll 1$, as long as $au \gg g'_s$ we can still trust
the NS5-branes solution. In this regime, setting
$au=e^{\Phi/\sqrt{Nb'}}$ we have:
\be
{ds'}^2=\frac{{l'}^2_s}{b'}[dx^2_{||}+d\Phi^2+Nb'd\Omega_3^2],\;\;\;\;g_s(\Phi)=
{g'}_se^{-Q\Phi}
\ee
where $Q=\frac{1}{\sqrt{Nb'}}$.  

Finally, for the case where $au \ll g'_s$, we have to
use the S-dual picture, which maps the theory to an ordinary commutative 
D5-branes.

Using the same variable as above, solution (\ref{NSBD}) can be written
as follows:
\bea
\frac{b}{{l'}^2_s}ds^2&=&k^{-1/2}(dx_4^2+dx_5^2)+k^{1/2}(dx_{||}^2+d\Phi^2
+Nb'd\Omega^2_3)\cr
g^2(\Phi)&=&{g'}^2_se^{-2Q\Phi}k,\;\;\;\;\;\;k=1+e^{2Q\Phi}.
\eea
From this form of the solution, one can see the deformation of the linear dilaton
background manifestly. The differential equation for the scalar in this
background, setting $\Psi=\psi e^{i\omega t}$, is:
\be
\partial_{\Phi}^2\psi+2Q\partial_{\Phi}\psi+\omega^2\psi=0.
\ee
This equation has a wave-like solution if $\omega > Q$. Actually one can
use this equation to study the absorption cross section of the polarized
graviton in this background. Doing so, we find \cite{AIO} that the absorption
cross section can be non-zero at the decoupling limit for an energy 
\be
\omega > \frac{{b'}^{-1/2}}{\sqrt{N}}.
\ee

We note that this is very similar to what was found in \cite{MS} for 
Type IIA NS5-branes. This relation, together with what we have had till now,
could suggest that $b'$ (or $b$) has a role of scale in 
non-commutative NS5-branes theory. From the smeared 
D3-branes point of view, this
can be considered as a scale that measures the smeared directions 

We can also calculate the Wilson loop (or 't Hooft loop) for the smeared
D3-branes that appeared above. Doing so we will find 
the same problem as ordinary
D5-branes \cite{BISY} namely that the distance $L$ between quark and
 antiquark
 does
not depend on $u_0$ (where $u_0$ is the minimal value of $u$). In fact we have
\be
L\sim \sqrt{b'}\sqrt{N}.
\ee 

Since this system is equivalent to a non-local theory, this classical minimal
distance should be related to the scale of non-locality and in fact this is the
case, except for the fact that it is larger by a factor of $\sqrt{N}$ than the
naively expected scale which was already mentioned in \cite{{MR},{HI}}.

Following \cite{HI2} we can also study the T-duality of the theory on the
D5-branes in the presence of a B field as well as non-commutative NS5-branes,
which has to be related to the Morita equivalent of the non-commutative theories
on the compact space.

\section{Non-commutative Type IIA NS5-branes}

{\large {\it Decoupling limit}}

The decoupling limit of the Type IIA NS5-branes can be obtained by T-duality from
the Type IIB one. Unlike the ordinary case, where the decoupling limit
for both IIB and IIA was the same, here we actually find that these limits
are different. The decoupling limit of the Type IIA NS5-brane in the 
presence of a non-zero RR field along its worldvolume is defined as
follows:
\be
l_s\rightarrow 0,\;\;\;\;\;\; l_s^3A={\rm fixed},\;\;\;\;\;\;\; g_sl^{-1}_s=
{\rm fixed}
\ee  
where $A$ is the RR field. As we see, this decoupling limit is completely 
different from the one we have in the ordinary NS5-branes and in fact it
is exactly the decoupling limit of smeared ordinary D2-branes. 
We will go back to this point later. The power of 3 in the fixed quantity
$l_s^3A$ can be understood by the fact that in Type IIA we are dealing with an
RR 3-form. We also note that this decoupling limit is different from the one
considered in \cite{AB}, where the same theory has been considered in the
DLCQ context; it is not clear to me whether these two definitions
are related or if they lead to two different deformations of the NS5-branes 
theory.
\footnote{I would like to thank O. Aharony for a discussion on this point.}

\vspace*{.4cm}

{\large {\it Supergravity description}}
\vspace*{.4cm} 

The supergravity solution of the Type IIA NS5-branes in the presence of a 
non-zero RR field is as follows:
\bea
ds&=&h^{-1/2}[-dx_{0}^2+dx^2_{1,2}+hdx_{3,4,5}^2+f(dr^2+r^2d\Omega_3^2)]\cr
&&\cr
f&=&1+\frac{ N {l}_s^2}{\cos \theta\; r^2},\;\;\;\;\
h^{-1}=\sin^2 \theta f^{-1}+\cos^2 \theta,\cr
&&\cr
e^{2\phi}&=&{g}_s^2f h^{-1/2},\;\;\;\;\;\;\;\;\;
 A_{345}=\frac{\tan\theta}{g_s}f^{-1}h \cr
A_{012}&=&\frac{\sin\theta}{g_s}f^{-1}.
\label{NSSA}
\eea

The decoupling limit is defined by taking the limit $l_s\rightarrow 0$ and 
 keeping fixed
\be\ba {ll}
u=\frac{r}{l_s^2}, &{\bar g}_s= g_sl_s^{-1}, \cr
b=l_s^2 \tan \theta, & {\bar x}_{3,4,5}=\frac{b}{l_s^2} x_{3,4,5} \ ,
\label{SCALA}
\ea\ee

From the NS5-branes point of view, we would expect, in the decoupling
limit, the mass of a D2-brane stretched between two NS5-branes to be fixed. 
If it had been the case, we would have had $\frac{r}{g_sl_s^3}$ fixed.
But here, in the decoupling limit, we have $\frac{r}{l_s^2}$ fixed, which
means that we are dealing with wrapped NS5-branes. On the other hand, this is
the only decoupling limit of NS5-branes that is consistent with string
theory dualities. Therefore what we are really
 studying in this section is non-commutative NS5-branes
theory wrapped on a circle. This can also be understood from the T-duality
we used to find the decoupling limit. As a result, beside the fixed quantities
defined in (\ref{SCALA}), we also have $\frac{R_s}{g_sl_s}$ fixed (here
$R_s$ is the radius of the compacted direction).
This extra condition will play an interesting role in the phase diagram
of the theory. Moreover, there is another parameter in the theory, which
we have to take into account: $\beta=g_s^2b$ \cite{AJO}.

In the decoupling limit the supergravity solution reads:
\bea
ds&=&\frac{l_s^2}{b}h^{1/2}\left[-dx_{0}^2+dx^2_{1,2}+h^{-1}dx_{3,4,5}^2+
\frac{Nb}{u^2}(du^2+u^2d\Omega_3^2)\right]\cr
&&\cr
h&=&1+(au)^2,\;\;\;\;\;\;a^2=\frac{b}{N},\;\;\;\;e^{2\phi}=
{\bar g}_s^2b\;\frac{\sqrt{1+(au)^2}}{(au)^2}\cr
&&\cr
A_{012}&=&\frac{l_s^3}{{\bar g}_sb^2}\;(au)^2,\;\;\;\;\;\;\;
A_{345}=\frac{l_s^3}{{\bar g}_sb^2}\;\frac{(au)^2}{1+(au)^2}.
\eea

An important point for this theory is that, in the limit where non-commutative
effects are not important and where, moreover, we have to lift the theory to 
M-theory,
the extra condition mentioned above means $\frac{R_s}{R_{11}}$ fixed, with
$R_{11}$ the radius of the 11th direction. This means that these circles have to
be in the same order or, in other words, that these two circles will 
decompactify
at the same energy. Therefore in the extreme IR we will end up with non-compact
(0,2) theory. On the other hand, in the extreme UV limit where the non-commutative
effects are important, $au\gg 1$, the solution reduces to:
\bea
l_s^{-2}ds&=&\frac{u}{R^2}(-dx_{0}^2+dx^2_{1,2})+
\frac{R^2}{u}(du^2+u^2d\Omega_3^2)+\frac{R^2}{b^2u}dx_{3,4,5}^2\cr
&&\cr
R^2&=&\sqrt{Nb},\;\;\;\;\;\;\;e^{2\phi}={\bar g}_s^2\;\frac{R^2}{u}.
\eea
Moreover, we have a non-zero 3-form along the directions (0,1,2) and (3,4,5),
the latter of which can be gauged away. Therefore we end up with an 
ordinary D2-branes
solution smeared in three directions\footnote{The curvature of the metric
is $l_s^2{\cal R}\sim \frac{1}{R^2 u}$; therefore one can trust the 
solution at UV. We note that this is in contrast with what we have
for non-smeared D2-branes where, at UV, we have perturbative SYM theory.}. 
Actually we could reach the same conclusion from Type IIB NS5-branes
as from using T-duality as we did.  

As the energy starts increasing from the IR limit, there are two possibilities.
If $\beta \gg 1$, first we reach a regime where we have the wrapped M5-branes
in the presence of a C field here the effects of the C field are important and
 we then have
to go to the Type IIA theory where the theory is described by smeared D2-branes.
On the other hand, if $\beta \ll 1$, we have to go to the Type IIA description,
where the non-commutative effects are not important and we can trust the
supergravity solution of NS5-branes; in fact, setting
$au=e^{\Phi/\sqrt{Nb}}$, we reach the same linear dilaton regime as in the
Type IIB case

\be
{ds}^2=\frac{{l}^2_s}{b}[dx^2_{||}+d\Phi^2+Nbd\Omega_3^2],\;\;\;\;g^2_s(\Phi)=
{\bar g}^2_sbe^{-2Q\Phi},
\ee
and finally we will reach at UV the regime that is described by smeared D2-branes.

As in the Type IIB case we can write the solution in terms of the new 
variable defined above
\bea
\frac{b}{{l'}^2_s}ds^2&=&k^{-1/2}(dx_3^2+dx_4^2+dx_5^2)+k^{1/2}(dx_{||}^2+d\Phi^2
+Nb'd\Omega^2_3)\cr
g^2(\Phi)&=&{\bar g}^{\prime 2}_sbe^{-2Q\Phi}k^{1/2},\;\;\;\;\;\;k=1+e^{2Q\Phi},
\eea
which shows the deformation of linear dilaton background in the presence of an
RR field.

One can also study this system using M5-branes in the presence on a non-zero
C field. Using the solution of M5-branes in the presence of a non-zero C field
\cite{{MR},{AJO}}, in the near-horizon region we have
\bea
ds^2&=&(NA)^{2/3}(1+\eta A^{-1})^{1/3}l_p^2\left[A^{-1}(dy_{0,1,2}^2+
\frac{dy_{3,4,5}^2}{1+\eta A^{-1}})+d\chi^2+d\rho^2+\rho^2d\Omega_3^2\right]\cr
&&\cr
A&=&\sum_{n=-\infty}^{n=\infty}\frac{\pi}{[\rho^2+(\chi-2\pi n)^2]^{3/2}},\;\;\;\;
\rho=\frac{r}{R_{11}},\;\;\;\;\;\;\chi=\frac{x_{11}}{R_{11}};
\eea
moreover, we rescaled the coordinates along the branes by $y_{0,1,2}=\sqrt{\frac{
\cos\theta}{l_s^2N}}\;x_{0,1,2}$ and $y_{3,4,5}=\sqrt{\frac{1}{l_s^2N\cos\theta}
}\;x_{3,4,5}$. The non-commutative NS5-branes solution is the regime where 
$\rho \gg 1$, while $\rho \ll 1$ is the M5-branes geometry. We note that
$\eta=\frac{R_{11}^2}{l_s^2N\cos\theta}$ is equal to $\frac{\beta}{N}$ at the
decoupling limit; as we can see from the solution for $\eta \ll 1$ we will
reach the linear dilaton regime before we go to the non-commutative NS5-branes
regime, but for $\eta \gg 1$ we will reach the non-commutative NS5-branes 
regime when we go to the Type IIA description.

This solution can be used in order to study the absorption cross section of 
the polarized graviton along the branes. Using the same notation as in \cite{MS},
the equation for minimally scalar of the form $\Psi=\psi e^{-i\sqrt{s}y_0}$
reads
\be
\partial_{\chi}^2\psi+\frac{1}{\rho^3}\partial_{\rho}\rho^3\partial_{\rho}\psi
+ \sum_{n=-\infty}^{n=\infty}\frac{s\pi}{[\rho^2+(\chi-2\pi n)^2]^{3/2}} \; 
\psi=0,
\ee
where $s=\frac{\omega^2 N l_s^2}{\cos\theta}$. From this notation one can use
the results of \cite{MS}; in particular we can compute the two-point
functions of the theory for the operator that couples to the 
graviton, which can be
interpreted as a component of the energy-momentum tensor of the theory. In fact
the results are the same but of course with our $s$ defined above. We also note
 that as the ordinary NS5-branes, the absorption cross section can be 
non-zero at the decoupling limit for the case $s>1$, or for the energy 
larger than $\omega > \frac{b^{-1/2}}{\sqrt{N}}$.    

This result is the same as for the ordinary case except, that $b$ plays 
the role of the
scale of the theory as expected. From this scale we would expect the
mass gap that could appear at $s=1$, as the ordinary NS5-branes, 
to be of the order of this scale;, however it 
is smaller by a factor of $\sqrt{N}$. Actually this is also the case in the
ordinary NS5-branes; while the scale of the theory is $m_s$, the mass gap
is of the order of $\frac{m_s}{\sqrt{N}}$. We also note that, as we saw in the
Type IIB case, there is a factor $\sqrt{N}$ difference between the scale of
non-locality, which we see from gravity calculation, and what is expected from
world-sheet calculation.

We would also like to note that the same phenomenon has been observed for the 
Coulomb branch of the ${\cal N}=4$ SYM theory in 4 dimensions, where the mass 
gap is smaller than expected, from a gauge theory point of view, 
by a factor of 
$\sqrt{g_{YM}^2N}$ \cite{BS}. This fact, together with our previous 
observation, might mean that there are some stringy effects that change 
the behaviour of the theory at strong coupling.

\section{Discussion}

We could consider NS5-branes in the presence of a non-zero RR field with 
higher rank. In this case we would find non-commutative NS5-branes in their
decoupling limit. These solutions have been given in \cite{AJO}. 
Here we were only interested in the case where, at UV, we can
trust the NS5-branes solution; in other words we are interested in the case
where, at UV, the theory is thought to be  a non-commutative version of 
the little string theory.
In fact
this is only the case when we have the smallest rank for the RR field. This can
be seen for example from the phase diagram of D5-branes with a higher-rank 
B field
\cite{AJO}. So in this paper we only considered the solution with smallest rank.

Another point that we observed is that the non-commutative little string 
theories can be described by ordinary branes smeared in some directions.
For example the non-commutative Type IIB NS5-branes  can be described
by the ordinary D3-branes smeared in two directions. If this is really the case,
it means that the theory must be equivalent to a non-commutative SYM theory with
16 supercharges in 6 dimensions. On the other hand we know that ordinary
SYM theory in 6 dimensions is not renormalizable; in other words, in order
to define the theory we need some more information at UV. A way to solve
this problem is to assume that the theory flows to a fixed point at UV where 
we
have the little string theory. Now, from our discussion we see that 
there is another
way to make the theory to be well-defined. In this way the theory flows to a
non-commutative theory, which can be considered as either non-commutative
6-dimensional SYM theory or non-commutative NS5-branes theory. So non-commutativity
gives us another way to make the theory well-defined. This is also the case
for the 5-dimensional theory, where there were two possibilities: either it flows
to a fixed point at UV ((0,2) theory) or it has a non-commutative 5-dimensional
description at UV.

We found that the decoupling limit of the Type IIA NS5-branes can be 
defined only when the theory is wrapped on a circle. This is also consistent 
with our knowledge of a possible non-commutative version of NS5-branes, 
since this theory is only defined in the DLCQ context, where one direction
is automatically wrapped on a circle \cite{{SW},{AB}}. Nevertheless, the theory
flows to the ordinary non-compact (0,2) theory at the IR limit.
  
We also note that the relation between non-commutative D5-branes, NS5-branes and
D3-branes smeared in two directions might give us some information about
the Coulomb branch of the ${\cal N}=4$ SYM theory in four dimensions
\cite{{KFT},{FGPW},{BS},{CR}}.\footnote{This issue has been suggested by
A. Giveon}

\begin{center}

{\large {\bf Acknowledgements}}
\end{center}

I would like to thank O. Aharony, A. Brandhuber, A. Giveon, A. Kumar and 
G. Mandal for useful discussions. I am also
indebted to Y. Oz for interesting discussions, which motivated me to study
this subject, as well as for his comments. The work was partially supported by 
H. Hofer.

\end{document}